\documentclass[11pt,twoside]{article}
\usepackage{diagrams}
\usepackage{atmp041}
\usepackage[dvips]{graphicx,color}
\input epsf.tex
\newtheorem{theorem}{Theorem}[section]

\usepackage{isolatin1}
\newtheorem{definition}[theorem]{Definition}
\newtheorem{example}[theorem]{Example}

\theoremstyle{remark}

\theoremstyle{proposition}
\newtheorem{proposition}[theorem]{Proposition}

\numberwithin{equation}{section}

\newcommand{\blankbox}[2]{\parbox{\columnwidth}{\centering}}
\input{diagrams.tex}
\newarrow{TTo}{<}{-}{-}{-}{>}
\newarrow{Line}{-}{-}{-}{-}{-}
\newarrow{Iline}{=}{=}{=}{=}{=}
\newarrow{Implies}{=}{=}{=}{=}{=>}
\newarrow{Dline}{}{dash}{}{dash}{}
\begin{document}
\title{\parbox[c]{13cm}{Quantum symmetries of face models}\\ and the double triangle algebra}
\author{Roberto Trinchero}
\address{Centro At\'omico Bariloche e Instituto Balseiro,
         8400 Bariloche, Argentina.}
\addressemail{trincher@cab.cnea.gov.ar}
\address{\small{The author is supported by CONICET.}}
\markboth{\it QUANTUM SYMMETRIES OF FACE MODELS\ldots}{\it R. TRINCHERO}
\begin{abstract}
Symmetries of trigonometric integrable two dimensional statistical face models
are considered. The corresponding symmetry operators on the Hilbert space of states of the quantum version of these models define a weak *-Hopf algebra isomorphic to the Ocneanu double triangle algebra.  
\end{abstract}
\cutpage

\section{Introduction}

This paper deals with the correspondence between rational conformal field theories of $SU(2)$-type and $ADE$ graphs\cite{z}.

The first and most basic relation between these two notions comes from the form 
of the modular invariant partition function associated to the above mentioned theories.
It turns out that the characters appearing in the partition function are labelled by the Coxeter numbers
of ADE graphs\cite{cappelli, CIZ, Pasquier}.

Next there is the Ocneanu  algebra of quantum symmetries\cite{Ocneanu:paths}\cite{Ocneanu:Bariloche} of the 
theory under consideration. The structure of this algebra can be given in terms of a graph, the Ocneanu graph of quantum symmetries.
Knowledge of this graph allows, without any further aid, to directly write the partition function in terms of characters. Furthermore other objects with clear physical interpretation can also be constructed from the information contained in this graph\cite{Coque:Qtetra, CoqueGil:ADE, robmar, pz, Gil:thesis}.

In addition there is the double triangle algebra\cite{Ocneanu:paths}.
This algebra is\cite{pz,ct} a weak Hopf algebra\cite{bs, nill, NikVainFiniteQG, NVdepth, NikshychVainerman} and the  algebra of quantum symmetries  can be obtained as the algebra describing the tensor category of the double triangle algebra representations associated with one of its product structures(the same name sometimes denotes the bialgebra itself). 
The elements of the double triangle algebra are certain endomorphisms of the vector space of paths over the corresponding ADE graph.
Thus by means of a purely mathematical construction\cite{ct} one can obtain the double triangle algebra associated to an ADE graph. Then the Ocneanu graph  can be obtained as describing the tensor category of  representations of this algebra. Finally from the graph   the partition functions can be obtained{\footnote{Note, however, that the historical path has been just the opposite.}}.

On the other hand there are the face models\cite{baxter} related to rational conformal field theories of $SU(2)$-type\cite{Pasquier}.
These classical statistical models are defined in terms of ADE graphs and they have second order transitions points where the physics can be described by the corresponding rational conformal field theory.  

In this paper we obtain a relation between the face models and the double triangle algebra. Indeed we show that
this algebra is the weak Hopf algebra of symmetries of the face model.
This algebra will be obtained from the action of its generators on the Hilbert space of states of the face model. These generators being defined as  
linear hermitian operators that commute with the corner transfer matrix of the face model for any horizontal length.
This, in  our opinion, gives a clear physical interpretation of the double triangle algebra that was lacking.
Furthermore it provides a derivation of the assumptions in ref. \cite{ct} out of natural
physical requirements.
In addition some interesting by-products appear as a result of this study. 
Among them, to show that the construction can be carried out without referring to essential paths and that their role is to provide a simple way to know the dimension of the double triangle algebra. Another by-product is the  possible extension of the formalism to bioriented tree graphs that are non-ADE. In addition the construction proves to be very useful for the calculation of connections on Ocneanu's cells. This is shown in appendix B for the case of the $A_n$ graphs.

The relation between all these approaches is schematically described by fig. 1.

\begin{figure}
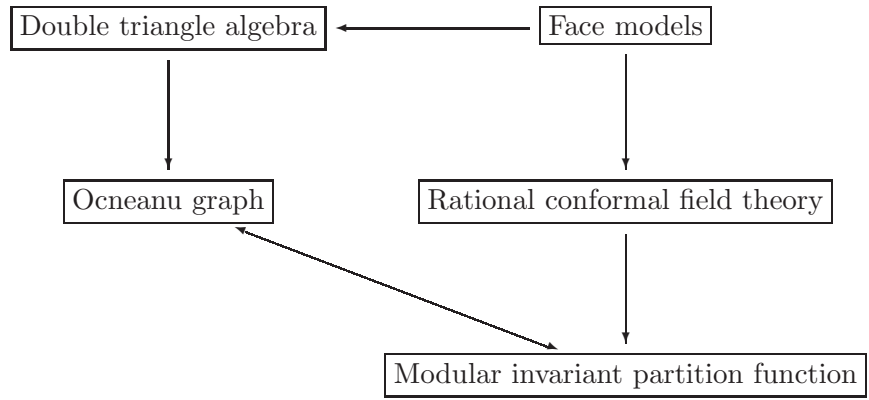

\label{0}
\begin{diagram}[size=3em]
\framebox{Double triangle algebra} & \lTo  &    \framebox{Face models}             \\
        \dTo   &       & \dTo                                                      \\
 \framebox{Ocneanu graph}&         & \framebox{Rational conformal field theory}    \\
                     &\rdTTo  &    \dTo                                             \\
       &        &              \framebox{Modular invariant partition function}     \\
\end{diagram}
\caption{Relationships}
\end{figure}

The paper is organised as follows. Section II  gives a very brief survey of integrable trigonometric face models. Section III defines what we mean by a symmetry transformation of the face models. Section IV describes certain consistency conditions related to the existence of the symmetry operators and studies the solutions to these conditions.
Section V describes the relation with Ocneanu cell calculus and Section VI 
makes contact with the double triangle algebra. The main sections are supplemented by two appendices.

\section{Face models: A brief survey}
\subsection{Variables and partition function\cite{baxter}}\label{fm}
Consider a lattice, such as the one drawn in fig. 2, with $N+1(M)$ horizontal(vertical) border vertices(in fig. 2 , $N =4 \;, M=3$). To each vertex in this lattice we associate
a vertex on a bioriented tree graph $G$ with $|V|$ vertices{\footnote{For some basic definitions and results on graph theory related to this work, see Appendix A.}}(such as the example given in fig. 3). This mapping is made 
in such a way that to nearest neighbours in the lattice{\footnote{Note that with the definition of a graph given in Appendix A, the lattice itself is a graph.}} we associate nearest neighbours
 in the graph{\footnote{If one defines a connected path in the lattice or in the graph as a succession of nearest neighbours, then this association corresponds to a continuous mapping from paths in the lattice to paths in the graph.}}.
The boundary conditions are chosen to be periodic in the vertical direction
and fixed in the horizontal direction
(the labelling of sites in fig. 2 is meant to show these boundary conditions).
We remark that we are interested not in one precise fixed horizontal boundary condition
but in the set of all possible ones.
To each elementary square in the lattice(such as the one with vertices $v_0 , v_1 , v_2 , v'_1$)we associate a weight $w \in {\mathbb R}$  that depends only on the corresponding mapping of vertices. We denote this number by $w(v_0 , v_1 , v_2 , v'_1)$.

\begin{figure}
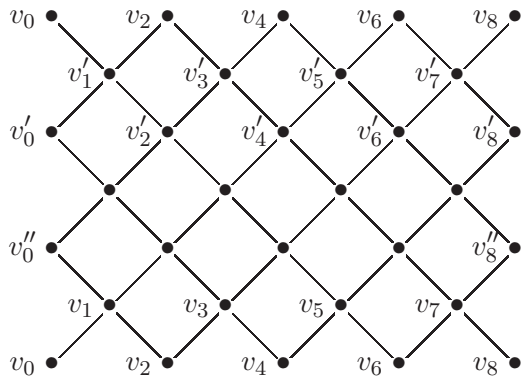

\label{1.1}
\begin{diagram}[size=1em]
v_0    & \bullet  &       &     & v_2    &\bullet &    &   & v_4 & \bullet    &       &     &    v_6 &\bullet&& &v_8&\bullet\\
 &       &\rdLine[abut]&       &\ldLine[abut]&      &\rdLine[abut]&      &\ldLine[abut]&       &\rdLine[abut]&       &\ldLine[abut]& &\rdLine[abut]&  &\ldLine[abut]&   \\
&       &    v'_1   &\bullet&       &      &    v'_3   &\bullet&       &       & v'_5      &\bullet&       &        &v'_7&\bullet     &&\\
&       &\ldLine[abut]&       &\rdLine[abut]&      & \ldLine[abut]& &      \rdLine[abut]&       &\ldLine[abut]&       &\rdLine[abut]& &\ldLine[abut]& &\rdLine[abut]&     \\
 v'_0   & \bullet  &       &       &  v'_2  &\bullet &        &       & v'_4 & \bullet    &       &       &   v'_6 &\bullet& & &v'_8& \bullet\\
 &       &\rdLine[abut]&       &\ldLine[abut]&      &\rdLine[abut]&      &\ldLine[abut]&       &\rdLine[abut]&       &\ldLine[abut]& &\rdLine[abut]& &\ldLine[abut]&     \\
&       &       &\bullet&       &      &        &\bullet&       &        &       &\bullet&       & &&\bullet&&\\
&       &\ldLine[abut]&       &\rdLine[abut]&      & \ldLine[abut]&  &\rdLine[abut]&       &\ldLine[abut]&      &\rdLine[abut]& &\ldLine[abut]& &\rdLine[abut]&\\   
v''_0&\bullet  &       &       &       &\bullet &        &       &       & \bullet   &       &       &       & \bullet&& &v''_8& \bullet \\
&       &\rdLine[abut]&       &\ldLine[abut]&      &\rdLine[abut]&   &\ldLine[abut]&       &\rdLine[abut]&       &\ldLine[abut]&    &\rdLine[abut]& &\ldLine[abut]& \\
&       &    v_1   &\bullet&       &      &   v_3     &\bullet&       &&     v_5   &\bullet&       &&v_7&\bullet&& \\
&       &\ldLine[abut]&       &\rdLine[abut]&      & \ldLine[abut]& &      \rdLine[abut]&       &\ldLine[abut]&       &\rdLine[abut]&   &\ldLine[abut]& &\rdLine[abut]& \\   
v_0 & \bullet  &       &       &     v_2  &\bullet &        &       &   v_4 & \bullet    &       &       &     v_6  &\bullet && &v_8&\bullet\\
\end{diagram}
\caption{Lattice for the face model.}
\end{figure}
\begin{figure}
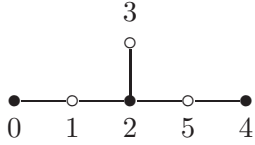

\label{1.2}
\begin{diagram}[size=1em]
      &             &       &      &3       &       &    &       &  \\
      &             &       &      &\circ       &       &    &       &  \\
      &             &       &      &\dLine[abut]  &       &    &       &  \\
\bullet&\rLine[abut]&\circ &\rLine[abut]&\bullet &\rLine[abut] &\circ&\rLine[abut] &\bullet\\
0      &             &1       &      &2       &       &5    &       & 4 \\
\end{diagram}
\caption{An example of a graph $G(=E_6)$.}
\end{figure}

The partition function corresponding to these  models is defined by,
\begin{equation}
\label{2.10}
Z= \sum_{\{v\}} \prod_{\diamond} w(\diamond )
\end{equation}
where the product is over all elementary squares appearing in the lattice  and the summation is over all possible 
assignations of graphs vertices to the lattice sites compatible with the rule given above. 
\subsection{Corner transfer matrix}
\label{ctm}
We consider a lattice, analog to the one in fig. 2 with $N+1$ horizontal border vertices 
. To each horizontal zig-zag line of length $2 N $, such as the lower border 
in fig. 2, we  associate a state
that we denote by,
\begin{equation}
\label{2.20}
|v_0 v_1 \cdots v_{2 N} > 
\end{equation}
 These states are in 1-1 correspondence with successions of nearest neighbours  of length $2N$. These successions will be called elementary paths and provide a preferred basis of a complex vector space
${\mathcal P}_{2N}${\footnote{The dimension $d$ of ${\mathcal P}_{N}$ can be obtained using the adjacency matrix $M$ of the graph $G$ as $d= \sum^{|V|}_{i,j=1} (M^{N})_{ij} $}}. The elements of this vector space being complex linear combinations of elementary paths of length $2N$.
We define the corner operators $U_i$ by,
\begin{eqnarray}
\label{2.30}
U_1 |v_0 v_1\cdots v_{2N}>&= &
\sum_{v'_1} w(v_0 , v_1 , v_2 , v'_1) 
|v_0 v'_1 \cdots v_{2N}  >\nonumber\\
U_2 |v_0 v_1 v_2\cdots v_{2N}>&= &
\sum_{v'_2} w(v_1 , v'_2 , v_3 , v_2) 
|v_0 v_1 v'_2 \cdots v_{2N}  >\nonumber\\
U_3|v_0 v_1 v_2 v_3 \cdots v_{2N}  >&= &
\sum_{v'_3} w(v_2 , v_3 , v_4 , v'_3) 
|v_0 v_1 v_2 v'_3 \cdots v_{2N}  >\nonumber\\
\vdots
\end{eqnarray} 
The partition function for a lattice with $M$ vertical border sites($M=3$ in \ref{1.1})
can be written as,
\begin{equation}
\label{2.50}
Z= Tr \left[ 
( U_1 U_3 \cdots U_{2N-1} U_2 U_4 \cdots U_{2N} )^M
\right]
\end{equation}

\subsection{Face  Yang Baxter equation(FYBE), trigonometric solutions of the FYBE and the Temperley-Lieb-Jones algebra(TLJA)}
From now on we will consider the case in which the weights depend on one real parameter 
$\theta \in {\mathbb R}$.
The face Yang-Baxter  equation(FYBE) in terms of corner operators is given by, \cite{baxter},
\begin{eqnarray}
\label{2.60}
U_{i+1}(\theta) U_i  (\theta + \phi) U_{i+1} (\phi) &=&
U_i(\theta) U_{i + 1} (\theta + \phi) U_i (\phi) \nonumber\\
U_i(\theta) U_j(\phi) & = & U_j(\phi) U_i(\theta) \;\;,\; |i-j| >1
\end{eqnarray}  
A solution of this equation is,
\begin{equation}
\label{2.80}
U_i(\theta)= 1 + f_{\gamma}(\theta) e_i
\end{equation}
where{\footnote{See Appendix A Def. 3.18 for the definition of $\beta$ and $\mu_{v_i}$.}},
\begin{equation}
\label{2.90}
f_{\gamma}(\theta) = \frac{\sin{\theta}}{ \sin{(\gamma - \theta)}} \qquad, \beta= 2 \cos{\gamma}
\end{equation}
and the action of the operators $e_i$ on the states is given by,
\begin{equation}
\label{2.100}
e_i |v_0 \cdots v_{i+1}\cdots>= \frac{1}{\beta}
\sum_{v'_{i+1}}\sqrt{\frac{\mu_{v_{i+1}}\mu_{v'_{i+1}}}{\mu_{v_i}\mu_{v_i+2}}}
\delta_{v_i v_{i+2}} |v_0 \cdots v'_{i+1} \cdots >
\end{equation}
These last equations define a trigonometric solution of the FYBE.

The $e_i$ operators generate the TLJA\cite{jones} defined by the following relations{\footnote{The hermitean conjugate in the formulae bellow is taken with respect to a scalar product where the basis of elementary paths is orthonormal\cite{gdhj}.}},
\begin{eqnarray}
\label{2.110}
e_i e_{i+1} e_i&= &\frac{1}{\beta^2} e_i \qquad , e_i e_j = e_j e_i \;\;,\; |i-j| >1\nonumber\\
e_i^2 = e_i\;\;, e_i^{\dagger}&= &e_i
\end{eqnarray}
related to the Jones projections $e_i$ are the Ocneanu creation and annihilation operators\cite{Ocneanu:paths} $c^{\dagger}_i$ and $c_i$ defined by,
\begin{eqnarray}
\label{2.120}
c_i |v_o \cdots v_i v_{i+1}\cdots> &= & \delta_{v_i v_{i+2}}\sqrt{\frac{\mu_{v_{i+1}}}{\mu_{v_i}}}
|v_o \cdots v_i {\check{v}}_{i+1}{\check v}_{i+2}\cdots>
 \nonumber\\
c^{\dagger}_i |v_o \cdots v_i v_{i+1}\cdots> &= &
\sum_{v'_{i+1}}\sqrt{\frac{\mu_{v'_{i+1}}}{\mu_{v_i}}}
|v_o \cdots v_i v'_{i+1}v_{i}v_{i+1}\cdots> \;\;\;,
\end{eqnarray}
where the $\check\, $ denotes omission. Indeed the projection $e_i$ is written in terms of the operators $c_i$ and $c^{\dagger}_i$ as,
\begin{equation}
\label{2.130}
e_i = \frac{1}{\beta} \; c^{\dagger}_i \, c_i
\end{equation}
From now on we shall refer to the models built up from the solutions (\ref{2.90}) of the FYBE as  trigonometric face models. 
\section{Symmetries of the trigonometric face models}
\label{sym}
\subsection{Conditions on symmetry generators}\label{cond}
We will be looking for symmetry generators in the general sense of linear hermitian operators that commute with the corner transfer matrix of the face model for any horizontal length and any fixed horizontal boundary condition.
These linear operators $T$ will act on the space of states of the system,  $T:{\mathcal P} \to {\mathcal P}$, were we have denoted by ${\mathcal P}$ the separable Hilbert space of paths of finite length on the bioriented tree graph $G$. 

A basis of ${\mathcal P}$ is given by $\{ \xi_i \}$, where $\xi_i$ are the elementary paths defined in subsection \ref{ctm}. The dual basis $\{ \xi^i \}$ in ${\mathcal P}^*$ is defined by
$(\xi_i,{\xi'}^j)= \delta^j_i$, where $(,)$ denotes the bilinear pairing 
$(,):{\mathcal P}\otimes {\mathcal P}^* \to {\mathbb C}$. A basis of the endomorphisms 
$End({\mathcal P})$ of 
${\mathcal P}$ is given by $\{\xi_i \otimes \xi^j \}$.
 
Each elementary path $\xi$ in ${\mathcal P}$ has an starting vertex $s(\xi)$ and an ending vertex $r(\xi)$. Let us denote by ${\mathcal P}_{\alpha_i \beta_i}$ the space of paths starting at the vertex $\alpha_i$ in $G$ and ending at vertex $\beta_i$.
Let us call $End({\mathcal P})_{\alpha\beta}: {\mathcal P}_{\alpha_i \beta_i} \to
{\mathcal P}_{\alpha_f \beta_f}\;, \alpha=(\alpha_i,\alpha_f) \;,\beta=(\beta_i,\beta_f)$ 
the subspace of $End({\mathcal P})$ of linear mappings from ${\mathcal P}_{\alpha_i \beta_i}$ to ${\mathcal P}_{\alpha_f \beta_f}$. We have that $End({\mathcal P})$
is the direct sum of the $End({\mathcal P})_{\alpha\beta}$, i.e.
$End({\mathcal P})= \bigoplus_{\alpha \beta} End({\mathcal P})_{\alpha\beta}$.

The action of an operator $T_{\alpha \beta} \in End({\mathcal P})_{\alpha\beta}$
is despited in fig. 4 
\begin{figure}[h]
\label{3.1}
\begin{center}
\input{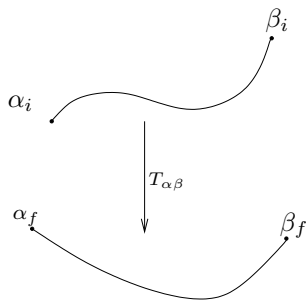}
\end{center}
\caption{The $T_{\alpha \beta}$ mapping.}
\end{figure}

For an operator \( T_{\alpha \beta } \) to be a symmetry operation
certain conditions will be required. Below we give the list of them
together with the corresponding physical interpretation. In this respect it is worth remarking that we are interested in the set{\footnote{We speak of a set because there are many trigonometric face models associated to the graph $G$ as in subsection \ref{fm},  differing one from the other by the fixed horizontal boundary conditions and  the number of horizontal and vertical border vertices.}} of trigonometric face models associated to a graph $G$ as in subsection \ref{fm}.

\begin{enumerate}
\item {\emph {Length should be preserved.}} That is the image of a length \( l \)
path should be a linear combination of length \( l \) paths. This
simply states that symmetry operations are endomorphism of the space
of states of models with the same horizontal length(that could differ one from the other by the horizontal boundary conditions). 
\item {\emph {Continuity.}}
In the space of paths ${\mathcal P}$ there is a natural product given by concatenation of paths. This product is defined for elementary paths and extends linearly to other paths. The concatenation product of two elementary paths is zero if the ending vertex of the first path is not equal to the starting vertex of the second  path. If the above holds then the product path is simply the extension of the first path by the second. In symbols
take $\xi_i = (v_0^i , \cdots , v_n^i)$ and 
$\xi_j = (v_0^j , \cdots , v_m^j)$ then the concatenation product $\xi_i \star \xi_j$ of $\xi_i$ and $\xi_j $ is given by, 
\begin{equation}
\label{3.10}
\xi_i \star \xi_j = \delta_{v_n^i v_0^j} 
(v_0^i , \cdots , v_n^i,v_1^j, \cdots , v_m^j) \qquad .
\end{equation}

We require that{\footnote{In eq.(\ref{3.20}) the terms in the summation over $\gamma=(\gamma_i, \gamma_f)$  will vanish if $\gamma_i \neq r(\xi)= s(\rho)$
or $\gamma_f$ is such that no path of length $\#\xi$ can join $\alpha_f$ to $\gamma_f$.}},
\begin{equation}
\label{3.20}
T_{\alpha \beta}(\xi \star \rho)= \sum_{\gamma} T_{\alpha \gamma}(\xi)\star 
T_{\gamma \beta}(\rho)
\end{equation}
this means that the mappings $T_{\alpha \beta}$ are continuous maps 
in the sense of the third footnote in section 2. From the physical point view it is clear that this must be fulfilled since a disconnected path makes no sense as a physical state.

\item {\emph {Hermiticity properties.}} We impose{\footnote{To show that this can always be done, suppose, $T^{\dagger}_{\alpha \beta}= T'_{{\tilde \alpha}{\tilde \beta}}$ then
define $T''_{\alpha \beta} = T_{\alpha \beta} + T'_{\alpha \beta}$ and 
${T'''}_{\alpha \beta} = i(T_{\alpha \beta} - T'_{\alpha \beta})$, both $T''$ and $T'''$
satisfy (\ref{3.30}). }},
\begin{equation}
\label{3.30}
T^{\dagger}_{\alpha \beta}= T_{{\tilde \alpha}{\tilde \beta}}
\end{equation}

\item {\emph {Preserved involution for the concatenation product.}}
Given a elementary path $\xi_i \in {\mathcal P}$ we can obtain another path $\xi_i^* \in {\mathcal P}$ reversing the sense in which the succession of vertices is followed. This operation followed by complex conjugation is an involution $*$ for the algebra of paths with respect to the concatenation product. We require this structure to be preserved by symmetry operations, that is,
\begin{equation}
\label{3.40}
T_{\alpha \beta}(\xi^* ) = (T_{\beta \alpha}(\xi))^*
\end{equation} 
\item {\emph {Preserve evolution.}}
 This corresponds to commutation of all the $T_{\alpha \beta}$ operators with the transfer matrix for a lattice of arbitrary horizontal length. This is achieved iff,
 \begin{equation}
 \label{3.50}
 [T_{\alpha \beta}, e_i]=0 \;\; \forall \alpha, \beta, i
 \end{equation} 
\end{enumerate}

\subsection{Conditions on the components of the symmetry generators}
Being a endomorphism of paths we can decompose  an operator $T_{\alpha \beta}$ in the basis $\{\xi_i \otimes \xi^j\}$ of $End({\mathcal P})$ as,
\begin{equation}
\label{3.60}
T_{\alpha \beta}= 
\sum_{i,j} 
{
\begin{diagram}[size=0.8em,abut]
\,&\xi_i  &\, \\
\bullet&\rTo~{n} & \bullet  \\
\dTo^{\alpha}&        &\dTo_{\beta} \\
\bullet&\rTo & \bullet  \\
\, &\xi^j  &\, \\
\end{diagram}
}\;
\xi_i \otimes \xi^j
= \sum_{\xi \xi'} 
{
\begin{diagram}[size=0.8em,abut]
\,&\xi  &\, \\
\bullet&\rTo~{n} & \bullet  \\
\dTo^{\alpha}&        &\dTo_{\beta} \\
\bullet&\rTo & \bullet  \\
\, &\xi'  &\, \\
\end{diagram}
}\;
\xi \otimes \xi'
\end{equation}
where the last equality is only a short hand notation, the prime over a elementary path indicating that it is the corresponding element in the dual basis in ${\mathcal P}^*$.
 The symbol,
\begin{equation}
\label{3.70}
{
\begin{diagram}[size=0.8em,abut]
\,&\xi  &\, \\
\bullet&\rTo~{n} & \bullet  \\
\dTo^{\alpha}&        &\dTo_{\beta} \\
\bullet&\rTo & \bullet  \\
\, &\xi'  &\, \\
\end{diagram}
}
\end{equation}
denotes a coefficient taking values in $\mathbb C$.
This symbol turns out to be useful to show the properties
of the $T_{\alpha \beta}$ operators. For example the definition of $T_{\alpha \beta}$
and condition 1. are summarised by saying that the symbol vanishes unless,
\begin{equation}
\label{3.80}
s(\xi)=\alpha_i \;,r(\xi)=\beta_i \;,s(\xi')=\alpha_f \;,r(\xi')=\beta_f\;, 
\#\xi'=\#\xi
\end{equation}
where the notation is that $\#\xi=n$ denotes the length of path $\xi$.
Condition 2. leads to,
\begin{equation}
\label{3.90}
{
\begin{diagram}[size=0.8em,abut]
\,&\xi_1  \star \xi_2   &\, \\
\bullet&\rTo~{n} & \bullet  \\
\dTo^{\alpha}&        &\dTo_{\beta} \\
\bullet&\rTo & \bullet  \\
\, &\xi'_1 \star  \xi'_2 &\, \\
\end{diagram}
}
=
{
\begin{diagram}[size=0.8em,abut]
\,&\xi_1  &\, \\
\bullet&\rTo~{n_1} & \bullet  \\
\dTo^{\alpha}&        &\dTo_{\gamma} \\
\bullet&\rTo & \bullet  \\
\, &\xi'_1  &\, \\
\end{diagram}
}
{
\begin{diagram}[size=0.8em,abut]
\,&\xi_2  &\, \\
\bullet&\rTo~{n_2} & \bullet  \\
\dTo^{\gamma}&        &\dTo_{\beta} \\
\bullet&\rTo & \bullet  \\
\, &\xi'_2  &\, \\
\end{diagram}
}
\end{equation}
where $n=n_1 + n_2$ and $\gamma$ is the pair of vertices $\gamma=(r(\xi_1),r(\xi'_1))$.
Condition 3.  in terms of coefficients is,
\begin{equation}
\label{3.100}
{
\begin{diagram}[size=0.8em,abut]
\,&\xi  &\, \\
\bullet&\rTo~{n} & \bullet  \\
\dTo^{\alpha}&       &\dTo_{\beta} \\
\bullet&\rTo & \bullet  \\
\, &\xi'  &\, \\
\end{diagram}
}=
{\overline
{
\begin{diagram}[size=0.8em,abut]
\,&\xi'  &\, \\
\bullet&\rTo~{n} & \bullet  \\
\dTo^{{\tilde \alpha}}&        &\dTo_{{\tilde \beta}} \\
\bullet&\rTo & \bullet  \\
\, &{\xi} &\, \\
\end{diagram}
}
}
\end{equation}
where ${\tilde \alpha}$ denotes the pair of vertices ${\tilde \alpha}=(\alpha_f ,\alpha_i)$. Condition 4. is,
\begin{equation}
\label{3.110}
{
\begin{diagram}[size=0.8em,abut]
\,&\xi  &\, \\
\bullet&\rTo~{n} & \bullet  \\
\dTo^{\alpha}&        &\dTo_{\beta} \\
\bullet&\rTo & \bullet  \\
\, &\xi'  &\, \\
\end{diagram}
}=
{\overline
{
\begin{diagram}[size=0.8em,abut]
\,&\xi^*  &\, \\
\bullet&\rTo~{n} & \bullet  \\
\dTo^{\beta}&        &\dTo_{\alpha} \\
\bullet&\rTo & \bullet  \\
\, &{\xi'}^* &\, \\
\end{diagram}
}
}
\end{equation}
Next we derive the consequences of condition 5.
In order to do so apply  (\ref{3.50}) for $i=0$  to a path $(v_0 v_1  v_2)$,
\begin{eqnarray}
\label{3.120}
T_{\alpha \beta} e_0 ((v_0 v_1 v_2) &=& \delta_{v_0 v_2}
\sum_{u_1} \sqrt{\frac{\mu_{v_1}\mu_{u_1}}{\mu^2_{v_0}}} T_{\alpha \beta} (v_0 u_1 v_0)
\nonumber\\
&=&
\delta_{v_0 v_2}
\sum_{u_1 v'_0 u'_1 v'_2} \sqrt{\frac{\mu_{v_1}\mu_{u_1}}{\mu^2_{v_0}}}
{
\begin{diagram}[size=0.8em,abut]
v'_0&u'_1  &v'_2 \\
\bullet&\rTo~{2} & \bullet  \\
\dTo^{\alpha}&        &\dTo_{\beta} \\
\bullet&\rTo & \bullet  \\
v_0 &u_1 &v_2 \\
\end{diagram}
}
(v'_0 u'_1 v'_2) 
\end{eqnarray}
on the other hand we have,
\begin{eqnarray}
\label{3.130}
e_0 T_{\alpha \beta}  ((v_0 v_1 v_2)) &=&
\sum_{v'_0 v'_1 v'_2} 
{
\begin{diagram}[size=0.8em,abut]
v'_0&v'_1  &v'_2 \\
\bullet&\rTo~{2} & \bullet  \\
\dTo^{\alpha}&        &\dTo_{\beta} \\
\bullet&\rTo & \bullet  \\
v_0 &v_1 &v_2 \\
\end{diagram}
}\,
e_0((v'_0 v'_1 v'_2))\nonumber\\
&=&
\delta_{v'_0 v'_2}
\sum_{u'_1 v'_0 v'_1 v'_2}
\sqrt{\frac{\mu_{u'_1}\mu_{v'_1}}{\mu^2_{v'_0}}} 
{
\begin{diagram}[size=0.8em,abut]
v'_0&v'_1  &v'_2 \\
\bullet&\rTo~{2} & \bullet  \\
\dTo^{\alpha}&        &\dTo_{\beta} \\
\bullet&\rTo & \bullet  \\
v_0 &v_1 &v_2 \\
\end{diagram}
}\,
(v'_0 u'_1 v'_2)
\end{eqnarray}
so (\ref{3.50}) implies,
\begin{equation}
\label{3.140}
\delta_{v_0 v_2}
\sum_{u_1}
 \sqrt{\frac{\mu_{v'_0}\mu_{u_1}}{\mu_{v_0} \mu_{u'_1}}}
{
\begin{diagram}[size=0.8em,abut]
v'_0&u'_1  &v'_2 \\
\bullet&\rTo~{2} & \bullet  \\
\dTo^{\alpha}&        &\dTo_{\beta} \\
\bullet&\rTo & \bullet  \\
v_0 &u_1 &v_0 \\
\end{diagram}
}
 =
\delta_{v'_0 v'_2}
\sum_{v'_1 }
\sqrt{\frac{\mu_{v_0}\mu_{v'_1}}{\mu_{v'_0}\mu_{v_1}}} 
{
\begin{diagram}[size=0.8em,abut]
v'_0&v'_1  &v'_0 \\
\bullet&\rTo~{2} & \bullet  \\
\dTo^{\alpha}&        &\dTo_{\beta} \\
\bullet&\rTo & \bullet  \\
v_0 &v_1 &v_2 \\
\end{diagram}
}\,
\end{equation}
which leads to,
\begin{equation}
\label{3.150}
\sum_{v'_1 }
\sqrt{\frac{\mu_{v_0}\mu_{v'_1}}{\mu_{v'_0}\mu_{v_1}}} 
{
\begin{diagram}[size=0.8em,abut]
v'_0&v'_1  &v'_0 \\
\bullet&\rTo~{2} & \bullet  \\
\dTo^{\alpha}&        &\dTo_{\beta} \\
\bullet&\rTo & \bullet  \\
v_0 &v_1 &v_2 \\
\end{diagram}
}\,
= \delta_{v_0 v_2} 
{
\begin{diagram}[size=1.3em,abut]
v'_0 \\
\dTo_{\alpha}\\
 v_0\\
\end{diagram}
}
=
 \delta_{v_0 v_2} 
{
\begin{diagram}[size=1.3em,abut]
v_0 \\
\dTo_{\alpha}\\
 v'_0\\
\end{diagram}
}
\end{equation}
where the symbols,
\begin{equation}
\label{3.160}
{
\begin{diagram}[size=1.3em,abut]
v_0 \\
\dTo_{\alpha}\\
 v'_0\\
\end{diagram}
}
\end{equation}
are coefficients taking non-negative real values, given by the l.h.s. of (\ref{3.150}) with $v_0 = v_2$.

We remark that one can apply condition (\ref{3.50}) to longer paths and consider other values of $i$. The relations so obtained can always be derived  employing (\ref{3.150}) and (\ref{3.90}).

\section{ Consistency equations}
\label{cons}
The next step is to study the solutions to conditions (\ref{3.80}), (\ref{3.90}), (\ref{3.100}), (\ref{3.110})
and (\ref{3.150}) of the previous subsection. It turns out that to have a solution certain consistency equations should be fulfilled by the coefficients (\ref{3.160}).
\subsection{Derivation}

In order to illustrate the methodology and the type of solutions to the 
conditions of the previous subsections we consider the following simple example.
\begin{example}\label{a3}
The case of $A_3 $.
The graph $G=A_3$ and its corresponding adjacency matrix $M$ are,
\begin{equation}
\label{4.10}
{
\begin{diagram}[size=0.8em,abut]
0 & \, & 1 & \, & 2 \\
\bullet&\rLine & \bullet &\rLine & \bullet \\
\end{diagram}
}
\qquad , \qquad
M=\left(
\begin{array}{lll}
 0&1 &0\\
 1&0 &1\\
 0&1 &0
 \end{array}
\right)
\end{equation}
where rows and columns are ordered as $0,1,2$ (the values for vertex $v$ can be $0,1$ or $2$).
The maximum eigenvalue is $\beta=\sqrt{2}$ and the Perron-Frobenius eigenvector is $(1,\sqrt{2},1)$.
Equations (\ref{3.150})) have a back-and-forth path for the upper horizontal side of the cell involved. Furthermore consider the case in which the r.h.s. of
(\ref{3.150}) is non-vanishing,i.e. $v_0 = v_2$. In this case the bottom horizontal side of the cell involved has also a back-and-forth length two horizontal path. Using (\ref{3.110}), each term in the l.h.s. of (\ref{3.150}) is the modulus square of a length one cell. For each site $\alpha_i$ and arbitrary $\alpha_f =v , v_1=v'$ nearest neighbours, you get one equation. The resulting three sets of equations(one equation in each set for each choice of nearest neighbours $v, v'$)  for this case are,
\begin{eqnarray}
\label{4.20}
\sqrt{
\frac{\mu_{1}\mu_{v}}{\mu_{v'}\mu_{0}}}\;\;
\left|{
\begin{diagram}[size=0.8em,abut]
0&  &1 \\
\bullet&\rTo~{1} & \bullet  \\
\dTo^{\alpha}&        &\dTo_{\beta} \\
\bullet&\rTo & \bullet  \\
v & &v' \\
\end{diagram}
}\right|^2
&= &
{
\begin{diagram}[size=1.3em,abut]
0 \\
\dTo_{\alpha}\\
 v\\
\end{diagram}
}
 \nonumber\\
\sqrt{\frac{\mu_{_1}\mu_{v}}{\mu_{v'}\mu_{2}}}\;\;
\left|{
\begin{diagram}[size=0.8em,abut]
2&  &1 \\
\bullet&\rTo~{1} & \bullet  \\
\dTo^{\alpha}&        &\dTo_{\beta} \\
\bullet&\rTo & \bullet  \\
v & &v' \\
\end{diagram}
}
\right|^2
&=& 
{
\begin{diagram}[size=1.3em,abut]
2 \\
\dTo_{\alpha}\\
 v\\
\end{diagram}
}
\nonumber\\
\sqrt{\frac{\mu_{0}\mu_{v'}}{\mu_{1}\mu_{v}}}\;\;
\left|{
\begin{diagram}[size=0.8em,abut]
1&  &0 \\
\bullet&\rTo~{1} & \bullet  \\
\dTo^{\alpha}&        &\dTo_{\beta} \\
\bullet&\rTo & \bullet  \\
v'& &v \\
\end{diagram}
}\right|^2
+
\sqrt{\frac{\mu_{2}\mu_{v'}}{\mu_{v}\mu_{1}}}\;\;
\left|{
\begin{diagram}[size=0.8em,abut]
1&  &2 \\
\bullet&\rTo~{1} & \bullet  \\
\dTo^{\alpha}&        &\dTo_{\beta} \\
\bullet&\rTo & \bullet  \\
v' & &v \\
\end{diagram}
}\right|^2
&=&
{
\begin{diagram}[size=1.3em,abut]
1 \\
\dTo_{\alpha}\\
 v'\\
\end{diagram}
}
\end{eqnarray}
where $vv'$ is a pair of nearest neighbours vertices in $G$.
The reflection properties (\ref{3.110}) imply,
\begin{equation}
\label{4.30}
\left|{
\begin{diagram}[size=0.8em,abut]
0&  &1 \\
\bullet&\rTo~{1} & \bullet  \\
\dTo^{\alpha}&        &\dTo_{\beta} \\
\bullet&\rTo & \bullet  \\
v & &v' \\
\end{diagram}
}\right|^2
=
\left|{
\begin{diagram}[size=0.8em,abut]
1&  &0 \\
\bullet&\rTo~{1} & \bullet  \\
\dTo^{\alpha}&        &\dTo_{\beta} \\
\bullet&\rTo & \bullet  \\
v'& &v \\
\end{diagram}
}\right|^2
\qquad,
\left|{
\begin{diagram}[size=0.8em,abut]
2&  &1 \\
\bullet&\rTo~{1} & \bullet  \\
\dTo^{\alpha}&        &\dTo_{\beta} \\
\bullet&\rTo & \bullet  \\
v & &v' \\
\end{diagram}
}
\right|^2
=
\left|{
\begin{diagram}[size=0.8em,abut]
1&  &2 \\
\bullet&\rTo~{1} & \bullet  \\
\dTo^{\alpha}&        &\dTo_{\beta} \\
\bullet&\rTo & \bullet  \\
v' & &v \\
\end{diagram}
}\right|^2
\end{equation}
using these last equations, one can replace the first two equations 
into the third equation in (\ref{4.20}) thus getting a consistency equation written 
purely in terms the coefficients (\ref{3.160}),
\begin{equation}
\label{4.40}
\frac{\mu_2}{\mu_v}
{
\begin{diagram}[size=1.3em,abut]
2 \\
\dTo\\
 v\\
\end{diagram}
}
-
\frac{\mu_1}{\mu_{v'}}
{
\begin{diagram}[size=1.3em,abut]
1 \\
\dTo\\
 v'\\
\end{diagram}
}
+
\frac{\mu_0}{\mu_v}
{
\begin{diagram}[size=1.3em,abut]
0 \\
\dTo\\
 v\\
\end{diagram}
}
=0
\end{equation} 
\end{example}
It is nice to find out that the consistency equation as in (\ref{4.40}) can be generalised for any bioriented tree graph.
This result
is given by the following theorem.
\begin{theorem}\label{cr}
Given a bioriented tree graph with $|V|$ vertices and any pair $vv'$ of nearest neighbours in $G$,
 the general form of the consistency equations, as 
(\ref{4.40}) for $A_3$, is given by,
\begin{equation}
\label{4.50}
\sum_{i=0}^{|V|-1} (-1)^{x_i}\frac{\mu_i}{\mu_{v(x_i)}}  
{
\begin{diagram}[size=1.3em,abut]
i \\
\dTo\\
 v(x_i)\\
\end{diagram}
}=0
\end{equation}
where $x_i$ is one of the two possible colours{\footnote{For the notion of colourability
and related results we refer the reader to appendix A}} $1$ or $0$, of the vertex $i=0,\cdots ,|V|-1$ and,
\begin{equation}
\label{4.60}
v(x_i)=\left\{ 
\begin{array}{l}
v \;{\rm if}\; x_i=0\\
v'\;{\rm if}\; x_i=1
\end{array}
\right.
\end{equation} 
\end{theorem}
\begin{proof}
We use induction in the number of edges $|A|$ in the graph. For $|A| =1$, the first value for which the condition makes sense, there 
is only one possible graph namely $A_2$, given by,
\begin{equation}
\label{4.70}
{
\begin{diagram}[size=0.8em,abut]
0 & \, & 1  \\
\bullet&\rLine & \circ  \\
\end{diagram}
}
\end{equation}
The consistency conditions (\ref{4.50}) are derived, as in the example of $A_3$, out of the equations obtained from (\ref{3.150}) with $\alpha_f = \beta_f$ for the graph in consideration. For the case of $|A|=1$(i.e., the case of the graph $A_2$) it is simple to check that they have the 
form (\ref{3.150}) with $|V| =2 $ and $\mu_0 = \mu_1 = 1$.

Now consider the case of a graph $G$ with $|A|$ edges. According to proposition
(\ref{nested}) in appendix A , there exists a family of subgraphs 
$G_a$ each one with $a$ edges $,\;a= 1, \cdots , |A|$ such that 
$G_1(=A_2 ) \subset G_2 \subset G_3 \cdots \subset G_{|A|}(=G)$.
Now suppose (\ref{4.50}) is valid for the graph $G_{|A|-1}$.
This equation is derived  as (\ref{4.40}) from (\ref{4.20}) in the example of $A_3$.
For given $v$ and $v'$, you have an equation for each vertex of the graph 
$G_{|A|-1}$, that involve the $\mu_i$ values of the corresponding $G_{|A|-1}$ graph.
The derivation of the consistency condition for the $G_{|A|}$ graph will follow the same steps as the one for the $G_{|A|-1}$ graph except for the following different features,
\begin{enumerate}
\item[(i)] The $\mu_i$ values are the ones corresponding to the graph $G_{|A|}$.
\item[(ii)] The equations (\ref{3.150}) with $\alpha_f = \beta_f$ that correspond to the vertex, call it $v_{|A|-1}$, where the additional edge of $G_{|A|}$ is inserted in
$G_{|A|-1}$, has an additional term. Denoting by $v_{|A|}$ the additional border vertex
that $G_{|A|}$ has with respect to $G_{|A|-1}$ the modified equation will be,
\begin{equation}
\label{4.80}
\sum_{w n.n. v_{|A|-1}} 
\sqrt{\frac{\mu_w \mu_v}{\mu_{v_{|A|-1}} \mu_{v'}}}
\left|{
\begin{diagram}[size=0.8em,abut]
v_{|A|-1}&  &w \\
\bullet&\rTo~{1} & \bullet  \\
\dTo&        &\dTo \\
\bullet&\rTo & \bullet  \\
v & &v' \\
\end{diagram}
}\right|^2
+
\sqrt{\frac{\mu_{v_{|A|}} \mu_v}{\mu_{v_{|A|-1}} \mu_{v'}}}
\left|{
\begin{diagram}[size=0.8em,abut]
v_{|A|-1}&  &v_{|A|} \\
\bullet&\rTo~{1} & \bullet  \\
\dTo&        &\dTo \\
\bullet&\rTo & \bullet  \\
v & &v' \\
\end{diagram}
}\right|^2
=
{
\begin{diagram}[size=1.3em,abut]
 v_{|A|-1}\\
\dTo \\
 v\\
\end{diagram}
}
\end{equation}
where the additional term is the last one in the l.h.s. of (\ref{4.80}).
\item[(iii)]There will appear an additional equation corresponding to the border vertex $v_{|A|}$ in  $G_{|A|}$, that does not belong to $G_{|A|-1}$ . This equation is,
\begin{equation}
\label{4.90}
\sqrt{\frac{\mu_{v_{|A|-1}} \mu_{v'}}{\mu_{v_{|A|}} \mu_{v}}}
\left|{
\begin{diagram}[size=0.8em,abut]
v_{|A|}&  &v_{|A|-1} \\
\bullet&\rTo~{1} & \bullet  \\
\dTo&        &\dTo \\
\bullet&\rTo & \bullet  \\
v' & &v \\
\end{diagram}
}\right|^2
=
{
\begin{diagram}[size=1.3em,abut]
 v_{|A|}\\
\dTo \\
 v'\\
\end{diagram}
}
\end{equation}
replacing (\ref{4.90}) in (\ref{4.80}), one gets,
\begin{equation}
\label{4.100}
\sum_{w n.n. v_{|A|-1}} 
\sqrt{\frac{\mu_w \mu_v}{\mu_{v_{|A|-1}} \mu_{v'}}}
\left|{
\begin{diagram}[size=0.8em,abut]
v_{|A|-1}&  &w \\
\bullet&\rTo~{1} & \bullet  \\
\dTo&        &\dTo \\
\bullet&\rTo & \bullet  \\
v & &v' \\
\end{diagram}
}\right|^2
=
{
\begin{diagram}[size=1.3em,abut]
 v_{|A|-1}\\
\dTo \\
 v\\
\end{diagram}
}
-
\frac{\mu_{v_{|A|}} \mu_v}{\mu_{v_{|A|-1}} \mu_{v'}}
{
\begin{diagram}[size=1.3em,abut]
 v_{|A|}\\
\dTo \\
 v'\\
\end{diagram}
}
\end{equation}
\end{enumerate}
that is the same as the equation for vertex $v_{|A|-1}$ of $G_{|A|-1}$ with the difference that the $\mu$'s are the ones of $G_{|A|}$ and the following replacement 
should be done,
\begin{equation}
\label{4.110}
{
\begin{diagram}[size=1.3em,abut]
 v_{|A|-1}\\
\dTo \\
 v\\
\end{diagram}
}
\to
{
\begin{diagram}[size=1.3em,abut]
 v_{|A|-1}\\
\dTo \\
 v\\
\end{diagram}
}
-
\frac{\mu_{v_{|A|}} \mu_v}{\mu_{v_{|A|-1}} \mu_{v'}}
{
\begin{diagram}[size=1.3em,abut]
 v_{|A|}\\
\dTo \\
 v'\\
\end{diagram}
}
\end{equation}
 So that what you will finally obtain as consistency condition will be the same as for $G_{|A|-1}$ but with the changes mentioned above. This leads exactly to (\ref{4.50}) for $G_{|A|}$.
\end{proof}

\subsection{Structure of the solutions to the consistency conditions}
Let us denote by ${\mathbb R}^+$ the non-negative real numbers. A solution $\omega$ of the consistency conditions (\ref{4.50}) is a set of numbers $\omega_{\alpha} \in {\mathbb R}^+$ for the variables (\ref{3.160}) for which (\ref{4.50}) holds. Since (\ref{4.50}) is a set of linear homogeneous equations for real and non-negative unknowns then any linear combination of solutions with coefficients in ${\mathbb R}^+$
is also a solution. This situation leads to the notion of purification of solutions.
\begin{definition}Purification of solutions and pure solution. A solution $\omega$ to (\ref{4.50}) can be purified iff there exists non-trivial linearly independent in ${\mathbb R}^+$ solutions $\omega^1$ and $\omega^2$ and non-vanishing numbers
$a_1, a_2 \in {\mathbb R}^+$ such that $\omega_{\alpha} = a_1 \omega^1_{\alpha} + a_2 \omega^2_{\alpha} \;\; \forall \alpha$. A solution is pure if it can not be further purified.
\end{definition}
The different pure solutions to the consistency equations (\ref{4.50}) will be labelled by an index $x$, the corresponding coefficients as in (\ref{3.70}) will have a label $x$
and the operators obtained from them as in (\ref{3.60}) will be denoted by $T^x_{\alpha \beta}$.

Regarding the indices $\alpha, \beta$ of the operators $T^x_{\alpha \beta}$ we have the following result relating them to the solutions of (\ref{4.50}).
\begin{proposition}\label{ind}
Given a solution $x$ to the consistency conditions (\ref{4.50}) the non-zero variables 
(\ref{3.160}) are in 1-1 correspondence with the indices $\alpha$ for which $T^x_{\alpha \beta}$ is non-vanishing for that solution and for some $\beta$.
\end{proposition}
\begin{proof}
Consider (\ref{3.150}) for the case $v_0 = v_2$. Using eq.(\ref{3.110}) it is clear
that in this case the l.h.s. of (\ref{3.150}) is a finite sum of moduli square. So
if the variable of the form  (\ref{3.160}) corresponding to the index $\alpha$ is non-vanishing them there is at least one index $\beta$ for which $T^x_{\alpha \beta}$ is non-vanishing
\end{proof}

\subsection{Solutions to the consistency conditions}

In this subsection we analyse  solutions of the consistency conditions
(\ref{4.50}). We have the following results,
\begin{proposition}\label{sol0}
For any connected finite bioriented tree graph there is a solution of (\ref{4.50}) given by,
\begin{equation}
\label{4.159}
{
\begin{diagram}[size=1.3em,abut]
 v_1\\
\dTo \\
 v_2\\
\end{diagram}
}
= 
\delta_{v_1 v_2 }
\end{equation}
\end{proposition}
\begin{proof}
Replacing (\ref{4.159}) in (\ref{4.50}) only two terms survive. With the notation of (\ref{4.50}) they are the ones where $i=v$ and $i=v'$. These terms have the same modulus and opposite signs.
\end{proof}

\begin{proposition}\label{sol1}
For any connected finite bioriented tree graph there is a solution of (\ref{4.50}) given by,
\begin{equation}
\label{4.160}
{
\begin{diagram}[size=1.3em,abut]
 v_1\\
\dTo \\
 v_2\\
\end{diagram}
}
= 
\left\{ 
\begin{array}{l}
1  \; if\; v_1 v_2 \;are \;nearest\;neighbours\\
0\; otherwise
\end{array}
\right.
\;\;.\end{equation}
\end{proposition}
\begin{proof}
Replacing (\ref{4.160}) in (\ref{4.50}) leads to,
\begin{equation}
\label{4.170}
-\sum_{v_i n.n. v} \frac{\mu_{v_i}}{\mu_v} +
\sum_{v'_i n.n. v'} \frac{\mu_{v'_i}}{\mu_{v'}} =0
\end{equation}
Next associate to $V$ a complex inner product vector space where the elements of $V$ are a orthonormal basis. Define the adjacency operator by,
\begin{equation}
\label{4.180}
M |v> = M_{v' v} |v'>
\end{equation}
where $M_{v'v}$ is as before the $v'v$ matrix element of the adjacency matrix.
Note that the operator $M$ is hermitian since $M_{v'v}$ are the matrix elements of a symmetric matrix.
Then we can write (\ref{4.170}) as,
\begin{equation}
\label{4.190}
- \frac{1}{\mu_v}<\mu| M v> + \frac{1}{\mu_{v'}}<\mu| M v'>=0
\end{equation}
where $\mu$ is the Perron-Frobenius eigenvector. Now since $M$ is hermitian and
$M |\mu >= \beta |\mu >$ by definition of the Perron-Frobenius eigenvector we get,
\begin{equation}
\label{4.200}
- \frac{1}{\mu_v}<\mu| M v> + \frac{1}{\mu_{v'}}<\mu| M v'>=
-\beta \frac{1}{\mu_v}<\mu|  v> + \beta \frac{1}{\mu_{v'}}<\mu|  v' >=0
\end{equation}
\end{proof}

Next we show the explicit solutions for the case of the graph $A_3$.
\begin{example}$A_3$. There are three linearly independent (in ${\mathbb R}^+$)
solutions, given by,
\begin{eqnarray}
\label{4.210}
{\mathbf 0} \qquad{
\begin{diagram}[size=1.3em,abut]
 0\\
\dTo \\
 0\\
\end{diagram}
}
=
{
\begin{diagram}[size=1.3em,abut]
 2\\
\dTo \\
 2\\
\end{diagram}
}
=
{
\begin{diagram}[size=1.3em,abut]
1\\
\dTo \\
1\\
\end{diagram}
}
=1 \qquad &&the \, rest\, zero. 
\nonumber\\
{\mathbf 1} \qquad{
\begin{diagram}[size=1.3em,abut]
 0\\
\dTo \\
 1\\
\end{diagram}
}
=
{
\begin{diagram}[size=1.3em,abut]
 1\\
\dTo \\
 0\\
\end{diagram}
}
=
{
\begin{diagram}[size=1.3em,abut]
2\\
\dTo \\
1\\
\end{diagram}
}
={
\begin{diagram}[size=1.3em,abut]
1\\
\dTo \\
2\\
\end{diagram}
}
=1 \qquad &&the \, rest\, zero. 
\nonumber\\
{\mathbf 2} \qquad{
\begin{diagram}[size=1.3em,abut]
 2\\
\dTo \\
 0\\
\end{diagram}
}
=
{
\begin{diagram}[size=1.3em,abut]
 0\\
\dTo \\
 2\\
\end{diagram}
}
=
{
\begin{diagram}[size=1.3em,abut]
1\\
\dTo \\
1\\
\end{diagram}
}
=1 \qquad &&the \, rest\, zero. 
\end{eqnarray}
Note that the first and second solutions in (\ref{4.210}) are the ones of propositions \ref{sol0} and \ref{sol1}
for this particular case.
\end{example}

Making a redefinition of connections(that will be given bellow)it is possible to show that for the solutions of the form (\ref{4.160})
eqs.(\ref{3.150}), (\ref{3.100}) and (\ref{3.110}) reduce to the  unitarity and reflection conditions  of ref. \cite{ct}. Thus those solutions corresponds to the fundamental irreps dealt with in that reference. That solutions are given explicitely for length 1 horizontal paths in appendix A of ref.\cite{ct} for the ADE graphs.  
In this respect it is useful to note that eqs. (\ref{3.150}) give information not only on the modulus  of the connections (information that is written explicitely for $A_3$ in
example \ref{a3}) but also on the phases. This information is given by equations (\ref{3.150}) when the r.h.s. vanishes and, although not proved here, this information is enough to completely determine, up to gauge equivalence,  the connections for the ADE graphs(see appendix A of ref. \cite{ct}).

It is also possible to build other symmetry operators satisfying the requirements of section \ref{sym}
by composing solutions of the consistency conditions (\ref{4.50}). 
However these new solutions are not of the type we have considered in sections 
\ref{sym} and \ref{cons}. They are not labelled by pair of vertices in $G$ but by more than two vertices in $G$. These solutions are dealt with in the following proposition,
\begin{proposition}
\label{comp}
The composition of operators satisfying conditions 1. to 5. of section \ref{cond}
also satisfy them.
\end{proposition}
\begin{proof}
Consider two operators $T^x_{\alpha \beta}$ and $T^y_{\gamma \delta}$ satisfying
conditions 1. to 5. of section \ref{cond}.
The composed operator $T^{x y}_{\alpha \cup \gamma \;\beta \cup \delta}= T^x_{\alpha \beta} \circ T^y_{\gamma \delta}$ obtained from the successive application of them to a path, can be written, in analogy with (\ref{3.60}), as,
\begin{equation}
\label{4.220}
T^{xy}_{\alpha \cup \gamma \;\beta \cup \delta}=
T^x_{\alpha \beta} \circ T^y_{\gamma \delta} = \sum_{\xi \xi'} \,
{
\begin{diagram}[size=0.8em,abut]
\,&\xi  &\, \\
\bullet&\rTo~{n} & \bullet  \\
\dTo^{\alpha \cup \gamma}&  xy      &\dTo_{\beta \cup \delta} \\
\bullet&\rTo & \bullet  \\
\, &\xi'  &\, \\
\end{diagram}
}\;
\xi \otimes \xi'
\end{equation}
where the connection involved turns out to be,
\begin{equation}
\label{4.230}
{
\begin{diagram}[size=0.8em,abut]
\,&\xi  &\, \\
\bullet&\rTo~{n} & \bullet  \\
\dTo^{\alpha \cup \gamma}&    xy    &\dTo_{\beta \cup \delta} \\
\bullet&\rTo & \bullet  \\
\, &\xi'  &\, \\
\end{diagram}
}\;
=
\delta_{\alpha_f \gamma_i}
\delta_{\beta_f \delta_i}
\sum_{\rho}
{
\begin{diagram}[size=0.8em,abut]
\,&\xi  &\, \\
\bullet&\rTo~{n} & \bullet  \\
\dTo^{\alpha}&   x     &\dTo_{\beta } \\
\bullet&\rTo & \bullet  \\
\, &\rho  &\, \\
\end{diagram}
}\;
{
\begin{diagram}[size=0.8em,abut]
\,&\rho  &\, \\
\bullet&\rTo~{n} & \bullet  \\
\dTo^{\gamma}&    y    &\dTo_{\delta} \\
\bullet&\rTo & \bullet  \\
\, &\xi'  &\, \\
\end{diagram}
}
\end{equation}
the indices $\alpha \cup \gamma$ denoting the triple of vertices $\alpha \cup \gamma=(\alpha_i \gamma_i \gamma_f)$ in $G$ and the same for $\beta \cup \delta$.

Now we consider the different properties that define a symmetry
operator as written in terms of connections in subsection 3.2.
It is clear that (\ref{3.90}) is satisfied by the composed connection(CC) appearing 
in the l.h.s. of (\ref{4.230}). Furthermore since the composition of continuous maps is a continuous map then (\ref{3.100}) also holds for the CC. It is also simple to verify  (\ref{3.110}) and (\ref{3.120}) for the CC. Finally the validity of (\ref{3.50})
for the composed operator is clear.
\end{proof}
How these new solutions are related to the ones that satisfy (\ref{3.150}) will be considered in the following sections.


\section{The relation with Ocneanu cell calculus} 

Replacing the solution (\ref{4.160}) in the relation (\ref{3.150}) we see that this equation is a condition on connections for cells with length one horizontal paths and indices $\alpha, \beta$ corresponding to nearest neighbours in the graph. Furthermore
defining new cells by,
\begin{equation}
\label{5.10}
{
\begin{diagram}[size=0.8em,abut]
v_0&\,  &v_1 \\
\bullet&\rTo~{n} & \bullet  \\
\dTo&  Oc      &\dTo \\
\bullet&\rTo & \bullet  \\
v_2 &\,  &v_3 \\
\end{diagram}
}
\,
=
\left(\frac{\mu_{v_1}\mu_{v_2}}{\mu_{v_0}\mu_{v_3}}\right)^{1/4}
\,
{
\begin{diagram}[size=0.8em,abut]
v_0&\,  &v_1 \\
\bullet&\rTo~{n} & \bullet  \\
\dTo&        &\dTo \\
\bullet&\rTo & \bullet  \\
v_2 &\,  &v_3 \\
\end{diagram}
}
\end{equation}
it is very simple to verify that the equations (\ref{3.150}), (\ref{3.110}) 
and (\ref{3.120}) of this paper correspond to the unitarity and reflection conditions for Ocneanu elementary connections\cite{Ocneanu:paragroups}\cite{Roche}\cite{EvansKawa:book} as written in eqs. (2.9) and (2.10) of ref. \cite{ct}.
Moreover (\ref{3.90}) remain the same in terms of the $Oc$ connections and is the same as (2.11) of ref. \cite{ct}, in addition eq. (2.7) of ref. \cite{ct} is the solution we dealt with in proposition \ref{sol0}. Finally the expression for the connection appearing in the l.h.s. of (\ref{4.230}) is the one associated to the tensor product representation 
that would follow from relation (2.12) of ref. \cite{ct} when written in terms of connections. 
These results are summarised in the following proposition,
\begin{proposition}
The connections $Oc$ defined by (\ref{5.10}) and corresponding to the solutions of the consistency conditions given in proposition \ref{sol1} satisfy conditions (2.7), (2.9), (2.10) and (2.11) of reference \cite{ct}.
In addition condition (2.12) of that reference, that defines the tensor product representation, when written in terms of connections coincides with eq.(\ref{4.230}). Thus the solutions in (\ref{5.10}) are identified with the so called fundamental connections of ref. \cite{ct}.
\end{proposition}


\section{The relation with the double triangle
weak Hopf algebra}
\subsection{The construction in $End({\mathcal P})$}
In order to make contact with the weak Hopf algebra of ref.\cite{ct}
we consider the dual space $End({\mathcal P})^*$ of linear forms on $End({\mathcal P})$.
A basis of $End({\mathcal P})$ is given by the elementary paths, i.e. 
$\{\xi_i \otimes \xi^j\} \;\; \xi_i \in {\mathcal P}, \;\xi^j\in {\mathcal P}^*$.
The dual basis $\{\xi^i \otimes \xi_j\}$ in $End({\mathcal P})^*$ is defined by,
\begin{equation}
\label{6.1}
(\xi_k \otimes \xi^l, \xi^i \otimes \xi_j) = \delta^i_k \delta^l_j
\end{equation}
where $(,):End({\mathcal P}) \otimes End({\mathcal P})^* \to {\mathbb C}$ is the bilinear pairing between both.

The symmetry operators satisfying conditions 1.-5. 
of section (\ref{sym}) span a finite dimensional subspace $A= \bigoplus_x A_x$ of the space $End({\mathcal P})$ of endomorphisms of Paths on $G$. The subspaces $A_x$ being the one 
spanned by the solution $x$ to the CC.

A basis of the subspaces $A_x$ is given by the $\{ T^x_{\alpha \beta} \}$, the dual basis 
$\{E_x^{\alpha \beta}\}$ in $(A^*_x)$ is defined by,
\begin{equation}
\label{6.2}
(T^x_{\alpha \beta}, E_x^{\gamma \delta})= \delta_{\alpha}^{\gamma}
\, \delta_{\beta}^{\delta}
\end{equation}
The elements $T^x_{\alpha \beta}$ of $End({\mathcal P})$
can be expressed in the $\{\xi_i \otimes \xi^j\}$
basis as,
\begin{equation}
\label{6.3}
T^x_{\alpha \beta} = \sum_{ij} ( T^x_{\alpha \beta}, \xi^i \otimes \xi_j ) \xi_i \otimes \xi^j
\end{equation}
which can be compared with (\ref{3.60}). Also in $End({\mathcal P})^*$ we have,
\begin{equation}
\label{6.4}
E^{\alpha \beta}_x = \sum_{ij} (\xi_i \otimes \xi^j , E_x^{\alpha \beta}) \xi^i \otimes \xi_j
\end{equation}
compatibility with (\ref{6.2}) implies,
\begin{equation}
\label{6.5}
\sum_{ij} ( T^x_{\alpha \beta}, \xi^i \otimes \xi_j )
(\xi_i \otimes \xi^j , E_x^{\gamma \delta})
= \delta_{\alpha}^{\gamma}
\, \delta_{\beta}^{\delta}
\end{equation}
A product $\cdot$ in $A^*$ directly related to the concatenation product $\star$ in 
$End({\mathcal P})$ will be defined. The point is that the concatenation product does not close in $A^*$. A projection is therefore employed,
\begin{equation}
\label{6.6}
E^{\alpha \beta}_x \cdot E^{\gamma \delta}_y =
P (E^{\alpha \beta}_x \star E^{\gamma \delta}_y)
\end{equation}
where the projector $P : End({\mathcal P})^* \to A^*$ is given by,
\begin{equation}
\label{6.7}
P  =\sum_z P_z \;\;,P_z =\sum_{\alpha \beta} E^{\alpha \beta}_z \otimes T^z_{\alpha \beta}
\;\; , P^2 = P 
\end{equation}
thus we have,
\begin{eqnarray}
\label{6.8}
E^{\alpha \beta}_x \cdot E^{\gamma \delta}_y &=&
P (E^{\alpha \beta}_x \star E^{\gamma \delta}_y)
 \nonumber\\
&= &\sum_z  P_z \sum_{ijkl}(\xi_i \otimes \xi^j , E_x^{\alpha \beta})
(\xi_k \otimes \xi^l , E_y^{\gamma \delta}) (\xi^i \otimes \xi_j) \star
(\xi^k \otimes \xi_l)
 \nonumber\\
&= & \sum_{z\eta \rho} E_z^{\eta \rho}
\sum_{ijkl}
(T^z_{\eta \rho},(\xi^i \otimes \xi_j) \star (\xi^k \otimes \xi_l))
(\xi_i \otimes \xi^j , E_x^{\alpha \beta})
(\xi_k \otimes \xi^l , E_y^{\gamma \delta})
 \nonumber\\
&= & \sum_{z\eta \rho} E_x^{\eta \rho}
\sum_{ijkl}
(T^z_{\eta \omega},\xi^i \otimes \xi_j)
(\xi_i \otimes \xi^j , E_x^{\alpha \beta})(T^z_{\omega\rho}, \xi^k \otimes \xi_l)
(\xi_k \otimes \xi^l , E_y^{\gamma \delta})
 \nonumber\\
&= & \sum_{z\eta \rho} E_x^{\eta \rho}
(T^z_{\eta \omega}, E_x^{\alpha \beta})(T^z_{\omega\rho}, E_y^{\gamma \delta})
 \nonumber\\
&= & \sum_{z\eta \rho} E_x^{\eta \rho}\delta_{zx}  \delta_{zy}
\delta^{\alpha}_{\eta} \delta^{\beta}_{\omega} \delta^{\gamma}_{\omega}
\delta^{\delta}_{\rho} = \delta_{xy}\delta^{\beta \gamma} E_x^{\alpha \delta}
\end{eqnarray} 
where we have used (\ref{6.4}) in witting the second equality, (\ref{6.7}) for the third equality, (\ref{3.90}) in the fourth, (\ref{6.5}) in the fifth and (\ref{6.2}) in the sixth. The above multiplication is matrix multiplication for the matrix units $E_x^{\alpha \delta}$'s.

This product gives $A^*$ the structure of a C$^*$-algebra.
Considering linear forms $\omega : A^* \to {\mathbb C}$ we can study representations of this C$^*$-algebra via the GNS construction\cite{haag}.
The positive definite normalised linear form $\omega_x : A_x^* \to {\mathbb C}$
associated to the pure solution $x$
defined by,
\begin{equation}
\label{6.9}
\omega_x ( E_y^{\alpha \beta} ) = \delta_{xy} \delta^{\alpha \beta} 
\end{equation}
is pure and therefore is associated to a irreducible representation of $A^*$\cite{haag}.
The algebra $A^*$ is generated by the $E_x^{\alpha \beta}$
with the product (\ref{6.8}). The scalar product is given by  the GNS definition,
\begin{equation}
\label{6.10}
 <E_x^{\alpha \beta}|E_x^{\gamma \delta}>
=\omega_x((E_x^{\alpha \beta})^* \cdot E_x^{\gamma \delta})=
\omega^x (E_x^{ \beta \alpha} \cdot E_x^{\gamma \delta})=
 \delta_{\alpha \gamma} \omega_x(E_x^{ \beta \delta})=
 \delta^{\alpha \gamma}\delta^{ \beta \delta}
\end{equation}
This can be repeated for all the pure solutions of the consistency conditions.
The direct sum of the representation spaces for each irreducible representation gives a total Hilbert space. From results related to the GNS construction we have the product law between objects in different irreps,
\begin{equation}
\label{6.11}
E^{\alpha \beta}_x \cdot E^{\gamma \delta}_y  = \delta_{xy} \delta_{\beta \gamma}
E^{\alpha \delta}_x
\end{equation} 
and the corresponding scalar product,
\begin{equation}
\label{6.12}
 <E_x^{\alpha \beta}|E_y^{\gamma \delta}>
= \delta_{xy} \delta^{\alpha \gamma}\delta^{ \beta \delta}
\end{equation}

Dual to the composition product in $End({\mathcal P})$ there exists a coproduct in 
$End({\mathcal P})^*$. This coproduct in the basis of elementary paths is  defined by,
\begin{equation}
\label{6.13}
((\xi_i \otimes \xi^j) \circ (\xi_k \otimes \xi^l) , \xi^m \otimes \xi_n )=
((\xi_i \otimes \xi^j) \otimes (\xi_k \otimes \xi^l), \Delta(\xi^m \otimes \xi_n))
\end{equation}
from which one obtains,
\begin{equation}
\label{6.14}
\Delta(\xi^m \otimes \xi_n) = \sum_p (\xi^m \otimes \xi_p) \otimes (\xi^p \otimes \xi_n) 
\end{equation}
This coproduct, being the dual of the composition product in $End({\mathcal P})$
, corresponds to the solutions considered in proposition \ref{comp}. In fact they correspond to the tensor product  representations of the algebra $A^*$.

Comparison  eqs.(\ref{6.4}), (\ref{6.5}), (\ref{6.10}), (\ref{6.11}) and (\ref{6.14})  
with (3.14), (3.15), (3.1), (3.5) and (4.3) of ref. \cite{ct} lead to the following theorem.
\begin{theorem}
The algebra $A^*$ is a weak Hopf algebra isomorphic to the one of ref. \cite{ct}.
\end{theorem} 

\subsection{The role of essential paths}

The subspace of essential paths ${\mathcal E}$ of the space of paths ${\mathcal P}$ on the graph $G$ is defined as follows.
\begin{definition}\label{ess}Essential subspace. It  is formed by the linear span of  paths $\xi$ such that,
\begin{equation}
e_i \, \xi = 0 \qquad , \forall i
\label{6.140}
\end{equation}
\end{definition}
Next the vector space of length preserving endomorphisms of essential paths
$End^{gr}({\mathcal E})$ is considered.
Recall that $A$ is the linear span of the operators 
$T_{\alpha \beta}$ and denote by $A_{\mathcal E}$ its restriction of  to $End({\mathcal E})$.
The following result is important.
\begin{proposition}
$End^{gr}({\mathcal E})$ and $A$ are isomorphic as vector spaces.
\end{proposition}
\begin{proof}
From proposition 3.3 of ref. \cite{ct} there is a 1-1 correspondence between 
$A_{\mathcal E}$ and $A$.
From eq. (\ref{3.50})  it is clear that $A_{\mathcal E} \subset End^{gr}({\mathcal E})$.
Next we show that all elements in $End^{gr}({\mathcal E})$ satisfy conditions 1.-5. of section \ref{sym}.

Condition 1. holds for any element in $End^{gr}({\mathcal E})$ by definition.

Condition 2. follows from the following argument.
Take  paths $\xi_i, \, \xi_m , \, \xi_n \in {\mathcal E}$ such that $\xi_i = \xi_m \star \xi_n$ 
(thus $\# \xi_i = \# \xi_m +  \# \xi_n$) and  paths 
$\xi_j , \,\xi_p ,\, \xi_q \in {\mathcal E}$ such that $\xi_j = \xi_p \star \xi_q$ 
(thus $\# \xi_j = \# \xi_p +  \# \xi_q$) and 
$\# \xi_j = \# \xi_i, \; \# \xi_m = \# \xi_p,\;\# \xi_n = \# \xi_q,$
then,
\begin{equation}
\label{6.150}
\xi_j \otimes \xi^i (\xi_i ) = \xi_p \otimes \xi^m (\xi_m ) \star 
\xi_q \otimes \xi^n (\xi_n )
\end{equation}

Condition 3. is the assertion that the corresponding set of operators is selfadjoint,
this follows from, 
\begin{equation}
\label{6.160}
(\xi_i \otimes \xi^j )^{\dagger} = \xi_j \otimes \xi^i 
\qquad \xi_i , \xi_j \in {\mathcal E} 
\end{equation}

In order to show  that condition 4. holds in  $End^{gr}({\mathcal E})$ first note that 
${\mathcal E}$ is closed under the involution involved in eq. (\ref{3.40}).
Furthermore $End^{gr}({\mathcal E})$ is closed under  
$* \otimes * : End^{gr}({\mathcal E}) \to End^{gr}({\mathcal E})$. Next note that,
\begin{equation}
\label{6.170}
\xi_k \otimes \xi^l  (\xi^*) = (\xi_k^* \otimes (\xi^l)^* (\xi))^*
\end{equation}
eq. (\ref{6.170}) is (\ref{3.40}).

The validity of condition 5. follows from ,
\begin{equation}
\label{6.180}
e_i (\xi_k \otimes \xi^l ) (\xi) = (\xi_k \otimes \xi^l ) e_i  (\xi) \qquad , \forall i
\end{equation}
with $\xi_k \otimes \xi^l \in End^{gr}({\mathcal E})$ and $\xi \in {\mathcal P}$.
Consider the l.h.s. of (\ref{6.180}). There are two possibilities either $\xi \in {\mathcal E}$ or not. If $\xi \in {\mathcal E}$ then its image by $\xi_k \otimes \xi^l$ would be in ${\mathcal E}$ and the application of 
$e_i$ will make the l.h.s. vanish. On the contrary if $\xi \notin {\mathcal E}$
the l.h.s. would also vanish because $\xi_k \otimes \xi^l \in End^{gr}({\mathcal E})$.
Now for the r.h.s., if $\xi$ is essential then it vanishes. If $\xi \notin {\mathcal E}$
then $P=e_i (\xi) \notin {\mathcal E}$ because $e_i (P)= e^2_i (\xi )= e_i (\xi)=P$, thus the image of $P$ by  $\xi_k \otimes \xi^l \in End^{gr}({\mathcal E})$ vanishes.
\end{proof}

\section*{Acknowledgements}
I would like to thank E. Andr\'es and S. Grillo for important suggestions related to this work. I am also indebted to R. Coquereaux and O. Ogievetsky for many enlightening talks on the subject and for constructive critical readings  of preliminary versions of the manuscript.  
\section*{Appendix A. Related graph theory.}
\begin{definition}[Finite (Oriented)Graph]
A finite (oriented)graph $G$ is a triple $G=(V, A, \varphi )$ where,
$V$ and $A$ are finite sets(whose elements are respectively called
vertices and edges) and $\varphi: A \to V \times V$ 
is a map that assigns to each edge in $A$
a (ordered)pair of vertices in $V$.
\end{definition}
Note that to every oriented graph $G$ it is possible to associate an unoriented one
$G_u=(V,A,\varphi_u)$ obtained from $G$ by disregarding the ordering of pairs in the image of $\varphi$.
\begin{definition}[Grade of a vertex]
The grade of a vertex in a graph $G$ is the number of edges in the associated unoriented
graph $G_u$ that contain this vertex
in their image by $\varphi_u$.
\end{definition}
\begin{definition}[Border vertex]
A border vertex $v \in V$ of a graph $G=(V, A, \varphi )$ is a vertex of grade one.
\end{definition}
We will be dealing with bioriented graphs.
\begin{definition}[Bioriented graph]
A graph $G=(V, A, \varphi )$ is bioriented iff $\varphi(a)= v_1 \times v_2, \;\; a \in A$ implies that there exists a edge $a^*$ such that $\varphi(a^*)= v_2 \times v_1$.
\end{definition}
\begin{definition}[Nearest neighbours]
A pair of vertices in a finite graph $G=(V, A, \varphi )$ are said to be nearest neighbours iff
they are the image by $\varphi$ of some edge in $A$. 
\end{definition}
\begin{definition}[Subgraph]\label{s}
A graph $G' = (V' ,A' , \varphi' )$ is said to be a subgraph of
a graph $G=(V, A, \varphi )$ iff,
\begin{enumerate}
\item[(i)] $V' \subset V$
\item[(ii)]$ A' \subset A$
\item[(iii)] $\varphi'$ is the restriction of 
$\varphi$ to $A'$
\end{enumerate}
\end{definition}
\begin{definition}[Path]
A path of length $n$ in a graph $G=(V, A, \varphi )$ is an alternated  succession of vertices and edges 
$(v_0, a_o , v_1, a_1 ,\cdots, a_{n-1} , v_n) \;\; v_i \in V \,,a_i \in A$ such that the vertex appearing before and after any edge are the image of that edge by $\varphi$. 
\end{definition}
\begin{definition}[Connected graph]
It is graph such that for every pair of vertices $v,v'$ there exists a path 
$\gamma=(v_0, a_o , v_1, a_1 ,\cdots, a_{n-1} , v_n)$ with $v_0 = v$ and $v_n = v'$ for some $n$.
\end{definition}
\begin{definition}[Cycle]
It is a path $(v_0, a_o , v_1, a_1 ,\cdots, a_{n-1} , v_n)$ such that $v_0 = v_n$
and all the vertices $(v_0, v_1, \cdots,  v_{n-1})$ are different
and $a^*_{n-1} \neq a_0$ .
\end{definition}
\begin{definition}[Tree graph]
It is a graph without cycles.
\end{definition}
Note that for a bioriented tree graph a pair of nearest neighbour vertices defines an edge uniquely. Thus a path in a bioriented tree graph can be determined by giving only a succession of nearest neighbours.
\begin{definition}[n-colourability]
A graph $G$ is n-colourable if n is the minimum number of colours required to color the vertices of $G$ in such a way that no nearest neighbours have the same color.
\end{definition}
Now we enunciate three propositions without proof.
\begin{proposition}\label{ep}
Any connected bioriented tree graph(CTG) has at least two border vertices.
\end{proposition}
\begin{proposition}\label{t}
Any connected subgraph of a CTG is a CTG.
\end{proposition}
\begin{proposition}
Any tree graph is 2-colourable.
\end{proposition}
\begin{definition}[Adjacency matrix]
One can characterise a bioriented tree graph $G$ by its adjacency matrix $M$. This matrix has size $|V| \times |V|$. Its $(v_1 , v_2)$ matrix element is
$1$ if vertex $v_1$ is connected  to vertex $v_2$, otherwise it vanishes. 
\end{definition}

\begin{definition}[Perron-Frobenius eigenvector]The normalised(set a smallest component to be equal to $1$) eigenvector with maximum eigenvalue $\beta$ of the adjacency matrix $M$ is called the Perron-Frobenius eigenvector
and its components will be denoted by $\mu_{v_i}, \;\;i=0, \cdots , |V|-1$.
\end{definition}

The following result is employed in the proof of theorem (\ref{cr})
\begin{proposition}\label{nested}
For every finite bioriented tree graph $G$ with $|A|$ edges there exists a family of subgraphs
$G_a \;\;, a= 1, \cdots , |A|$ and $G_{|A|}=G$ such that $G_a$ has $a$ edges and,
$G_1 \subset G_2 \subset G_3 \cdots \subset G_{|A|}$.
\end{proposition}
\begin{proof}
Due to prop. (\ref{ep}) $G$ has at least two endpoints.
Consider a subgraph $G_{|A|-1}$ obtained from $G$ by eliminating one of its endpoints and the
corresponding edge. Due to def. (\ref{s}) $G_{|A|-1}$ will be a subgraph of $G_{|A|}$. 
 Furthermore due to prop. (\ref{t}) this is also a bioriented tree graph.
 Repeat this procedure with $G_{|A|-1}$ to obtain $G_{|A|-2}$ and so on. Since $G$ is finite
this procedure will end up giving the required family of subgraphs.
\end{proof}
\newpage

\section*{Appendix B. The value of connections for $A_n$ graphs.}
The aim of this appendix is to provide the value of connections for the $A_n$
graphs for any of the irreps of the corresponding double triangle algebras. To do so we first study the solutions to the consistency conditions, then we give a general formula for the modulus of connections with length one horizontal paths for each of the above mentioned solutions and finally we give a choice of phases for these connections. These results will be presented in three propositions. Outlines for their proof are provided.

The consistency equations for the graphs $A_n$ for $n$ odd, are,
\begin{equation}
\label{b1}
\frac{\mu_{n-1}}{\mu_v}
{
\begin{diagram}[size=1.3em,abut]
n-1 \\
\dTo\\
 v\\
\end{diagram}
}
-
\frac{\mu_{n-2}}{\mu_{v'}}
{
\begin{diagram}[size=1.3em,abut]
n-2 \\
\dTo\\
 v'\\
\end{diagram}
}
\cdots
+
\frac{\mu_2}{\mu_v}
{
\begin{diagram}[size=1.3em,abut]
2 \\
\dTo\\
 v\\
\end{diagram}
}
-
\frac{\mu_1}{\mu_{v'}}
{
\begin{diagram}[size=1.3em,abut]
1 \\
\dTo\\
 v'\\
\end{diagram}
}
+
\frac{\mu_0}{\mu_v}
{
\begin{diagram}[size=1.3em,abut]
0 \\
\dTo\\
 v\\
\end{diagram}
}
=0
\end{equation}
 For $n$ even,
\begin{equation}
\label{b2}
\frac{\mu_{n-1}}{\mu_v}
{
\begin{diagram}[size=1.3em,abut]
n-1 \\
\dTo\\
 v\\
\end{diagram}
}
-
\frac{\mu_{n-2}}{\mu_{v'}}
{
\begin{diagram}[size=1.3em,abut]
n-2 \\
\dTo\\
 v'\\
\end{diagram}
}
\cdots
-
\frac{\mu_2}{\mu_{v'}}
{
\begin{diagram}[size=1.3em,abut]
2 \\
\dTo\\
 v'\\
\end{diagram}
}
+
\frac{\mu_1}{\mu_{v}}
{
\begin{diagram}[size=1.3em,abut]
1 \\
\dTo\\
 v\\
\end{diagram}
}
-
\frac{\mu_0}{\mu_{v'}}
{
\begin{diagram}[size=1.3em,abut]
0 \\
\dTo\\
 v'\\
\end{diagram}
}
=0 
\end{equation}
where  $v,v'$ can be any pair of nearest neighbours vertices in $A_n$.

The components $\mu_j$ of the Perron-Frobenius eigenvector for $A_n$ are given by,
\begin{equation}
\label{b3}
\mu_j = [j+1]_q = \frac{q^{j+1} - q^{-j-1}}{q - q^{-1}} = 
\frac{\sin{\frac{\pi (j+1)}{(n+1)}}}{\sin{\frac{\pi}{(n+1)}}}
 \;\;, q=\exp^{i \pi / (n+1)}
\end{equation}
Regarding the pure and normalised solutions to eqs.(\ref{b1})-(\ref{b2}) we have the following result.
\begin{proposition}
For each $l=0,1,\cdots , n-1$ there is a pure and normalised solution to eqs.(\ref{b1})-(\ref{b2}) for $A_n$  given by,
\begin{equation}
\label{b4}
{
\begin{diagram}[size=1.3em,abut]
 v_1\\
\dTo^l \\
 v_2\\
\end{diagram}
}
=1 \;\; if \; v_1 \; can\; be\; connected\; to\; v_2\; by\; a\; path\; of\; length\; l
\; on \; A_n\end{equation}
except for the cases,
\begin{equation}
\label{b5}
{
\begin{diagram}[size=1.3em,abut]
 p-2\\
\dTo^l \\
 l-p\\
\end{diagram}
}=0=
{
\begin{diagram}[size=1.3em,abut]
 n-1-(p-2)\\
\dTo^l \\
 n-1-(l-p)\\
\end{diagram}
}\;\; , p=2,3, 4, \cdots , n-1
\end{equation}
whenever the values of the vertex indices appearing in (\ref{b5}) make sense for the graph $A_n$.
\end{proposition}
This last assertion can be verified by replacing the solutions (\ref{b4})-(\ref{b5}) in eq.(\ref{b1})-(\ref{b2}).

Regarding the value of the modulus of cells with length one horizontal paths 
the following result holds.
\begin{proposition}
\begin{equation}
\label{b6}
\sqrt{\frac{\mu_{p}\mu_{p-1}}{\mu_{v'}\mu_{v}}}\;
\left| 
{
\begin{diagram}[size=0.8em,abut]
p-1&\,  &p \\
\bullet&\rTo~{1} & \bullet  \\
\dTo& l       &\dTo \\
\bullet&\rTo & \bullet  \\
v' &\,  &v \\
\end{diagram}
}
\right|^2
=
\frac{\mu_{p-1}}{\mu_{v'}}
{
\begin{diagram}[size=1.3em,abut]
 p-1\\
\dTo^l \\
 v'\\
\end{diagram}
}
-
\frac{\mu_{p-2}}{\mu_{v}}
{
\begin{diagram}[size=1.3em,abut]
 p-2\\
\dTo^l \\
 v\\
\end{diagram}
}
+
 \cdots
-(+)
\frac{\mu_{0}}{\mu_{v(v')}}
{
\begin{diagram}[size=1.3em,abut]
 0\\
\dTo^l \\
 v(v')\\
\end{diagram}
}
\end{equation}
where $v, v'$ are any pair of nearest neighbours in $A_n$,  $p= 1, \cdots , n-1$ and the values in parenthesis holding for the case  $p$ even and the others for $p$ odd. 
\end{proposition}
\noindent Eq.(\ref{b6}) is obtained in the same way as in the case of $A_3$, eq.(\ref{4.20}).

Regarding the choice of phases for these cells we have.
\begin{proposition}
For any $A_n$ the phases can be chosen to be $+1$ or $-1$. For the cells involved in the previous proposition we have,
\begin{equation}
\label{b7}
Phase \left(
{
\begin{diagram}[size=0.8em,abut]
p&\,  &p+1 \\
\bullet&\rTo~{1} & \bullet  \\
\dTo& l       &\dTo \\
\bullet&\rTo & \bullet  \\
v &\,  &v + 1 \\
\end{diagram}
}
  \right)= 1
  \;\;\;\;,
Phase \left(
{
\begin{diagram}[size=0.8em,abut]
p-1&\,  &p \\
\bullet&\rTo~{1} & \bullet  \\
\dTo& l       &\dTo \\
\bullet&\rTo & \bullet  \\
v &\,  &v - 1 \\
\end{diagram}
}
  \right)=  (-1)^{p-1}
\end{equation}
where $v= p ,p\pm 2, \cdots , p\pm l $ for $l$ even and
$v= p\pm 1 ,p\pm 3, \cdots , p\pm l $ for $l$ odd. Eqs. (\ref{b6}) are valid 
for $p$ such that the  vertex indices  appearing in (\ref{b4}) make sense for the graph $A_n$.
\end{proposition}
The first assertion is obtained using the general form of $A_n$ connections appearing in appendix A of ref. \cite{ct} and performing adequate gauge transformations.
The second assertion follows from replacing (\ref{b7}) in eq.(\ref{3.150}) for
$v_0 \neq v_2$.

\end{document}